%% file: Dstar0_branching_fractions.tex
\RequirePackage{lineno}
\documentclass[prd,twocolumn,showpacs,amsmath,amssymb,nofootinbib]{revtex4-1}
\usepackage{graphicx}
\usepackage{dcolumn}
\usepackage{subfigure}
\usepackage{epsfig}
\usepackage{float}
\usepackage{times}
\usepackage{setspace}
\usepackage{xcolor}

\newcommand{\BR}{{\cal B}}

\begin{document}
\normalsize
\parskip=5pt plus 1pt minus 1pt
\title{\boldmath Precision measurement of the $D^{*0}$ decay
branching fractions}

\input{authors_dec2014}

\begin{abstract}

Using 482~pb$^{-1}$ of data taken at $\sqrt{s}=4.009$~GeV, we measure
the branching fractions of the decays of $D^{*0}$ into
$D^0\pi^0$ and $D^0\gamma$ to be $\BR(D^{*0}
\to D^0\pi^0)=(65.5\pm 0.8\pm 0.5)\%$ and $\BR(D^{*0} \to
D^0\gamma)=(34.5\pm 0.8\pm 0.5)\%$ respectively, by assuming that the $D^{*0}$ decays only into these two modes.
The ratio of the two branching fractions is $\BR(D^{*0} \to D^0\pi^0)/\BR(D^{*0} \to D^0\gamma) =1.90\pm 0.07\pm 0.05$, which is independent of the
assumption made above. The first uncertainties are
statistical and the second ones systematic. The precision is
improved by a factor of three compared to the present world
average values.

\end{abstract}
\maketitle

\section{Introduction}

Quantum chromodynamics~(QCD)~\cite{Fritzsch:1973pi} is widely accepted
as the correct theory for the strong interaction. In the framework of
QCD, the building blocks of matter, colored quarks, interact with each other by
exchanging $SU(3)$ Yang-Mills gauge bosons, gluons, which are also
colored. Consequently, the quark-gluon dynamics becomes
nonperturbative in the low energy regime. Many effective models~(EMs), such as the potential model, heavy quark and chiral
symmetries, and QCD sum rules, have been developed to deal with the
nonperturbative effects, as described in a recent
review~\cite{review}. The charmed meson, described as a hydrogen-like hadronic
system consisting of a heavy quark~($c$ quark) and a light
quark~($u$, $d$, or $s$ quark), is a particularly suited laboratory to test the
EMs mentioned above. The decay branching fractions of $D^{*0}$ to
$D^{0}\pi^{0}$~(hadronic decay) and $D^{0}\gamma$~(radiative decay)
have been studied by a number of authors based on
EMs~\cite{Eichten:1979ms, Cheng:1993gc, Aliev:1994nq,
Miller:1988tz}. A precise measurement of the branching fractions will
constrain the model parameters and thereby help to improve the
EMs. On the experimental side, these two branching fractions are
critical input values for many measurements such as the open charm
cross section in $e^+e^-$ annihilation~\cite{CroninHennessy:2008yi}
and the semileptonic decays of $B^{\pm}$~\cite{Bozek:2010xy}.

These branching fractions have been measured in many electron-positron
collision experiments, such as CLEO~\cite{cleo}, ARGUS~\cite{argus},
BABAR~\cite{babar} etc., but the uncertainties of the averaged
branching fractions by the Particle Data Group~(PDG)~\cite{pdg} are
large~(about 8\%).  The data sample used in this analysis of
482~pb$^{-1}$ collected at a center-of-mass~(CM) energy $\sqrt{s} =
4.009$~GeV with the BESIII detector provides an opportunity for significant
improvement.

\section{BESIII detector and Monte Carlo}

BESIII is a general purpose detector which covers 93\%
of the solid angle, and operates at the $e^+ e^-$ collider BEPCII.
Its construction is described in great detail in Ref.~\cite{BESIII}.
It consists of four main components: (a) A small-cell, helium-based
main drift chamber (MDC) with 43 layers providing an average
single-hit resolution of 135~$\mu$m, and a momentum resolution of
0.5\% for charged-particle at 1~GeV/$c$ in a 1~T magnetic field. (b)
An electro-magnetic calorimeter (EMC) consisting of 6240 CsI(Tl)
crystals in a cylindrical structure (barrel and two end-caps). The
energy resolution for 1~GeV photons is 2.5\% (5\%) in the barrel
(end-caps), while the position resolution is 6~mm (9~mm) in the barrel
(end-caps). (c) A time-of-fight system (TOF), which is constructed of
5-cm-thick plastic scintillators and includes 88 detectors of 2.4~m
length in two layers in the barrel and 96 fan-shaped detectors in the
end-caps. The barrel (end-cap) time resolution of 80~ps (110~ps)
provides 2$\sigma$ $K/\pi$ separation for momenta up to about
1~GeV/$c$. (d) The muon counter (MUC), consisting of Resistive Plate
Chambers (RPCs) in nine barrel and eight end-cap layers, is
incorporated in the return iron of the super-conducting magnet,
and provides a position resolution of about 2~cm.

To investigate the event selection criteria, calculate the selection
efficiency, and estimate the background, Monte Carlo~(MC) simulated
samples including 1,000,000 signal MC events and 500~pb$^{-1}$ inclusive MC events are
generated. The event generator {\sc kkmc}~\cite{Jadach:1999vf} is used
to generate the charmonium state including initial state
radiation~(ISR) and the beam energy spread; {\sc
evtgen}~\cite{Ping:2008zz} is used to generate the charmonium
decays with known branching ratios~\cite{pdg}; the unknown
charmonium decays are generated based on the {\sc lundcharm}
model~\cite{Chen:2000tv}; and continuum events are generated with {\sc
pythia}~\cite{Sjostrand:2006za}. In simulating the ISR events, the
$e^{+} e^{-} \to D^{*0}\bar{D}^{0}$ cross section measured with
BESIII data at CM energies from threshold to 4.009~GeV is used as
input.  A {\sc geant4}~\cite{geant41,geant42} based detector
simulation package is used to model the detector response.

\section{Methodology and event selection}
At $\sqrt{s} = 4.009$~GeV, $e^+e^- \to D^{*0}\bar{D}^0+c.c.$ is
produced copiously. Assuming that there are only two decay modes for $D^{*0}$, i.e., $D^{*0}\to
D^{0}\pi^{0}$ and $D^{*0}\to D^{0}\gamma$, the final states of $D^{*0}\bar{D}^0$
decays will be either $D^0\bar{D}^0\pi^0$ or $D^0\bar{D}^0\gamma$.
Such an assumption is reasonable, since as shown in Ref.~\cite{next}, the next largest
branching fraction mode $D^{*0}\to D^{0}\gamma\gamma$ is expected to be less than $3.3\times 10^{-5}$.
The CM energy is not high
enough for $D^{*0}\bar{D}^{*0}$ production. To select $e^+e^- \to D^{*0}\bar{D}^0$ signal events, we first
reconstruct the $D^0\bar{D}^0$ pair, and then require that the mass
recoiling against the $D^{0}\bar{D}^0$ system corresponds to a $\pi^{0}$
at its nominal mass~\cite{pdg} or a photon with a mass of zero. This approach
allows us to measure the $D^{*0}$ decay branching ratios from the
numbers of $D^{*0} \to D^{0}\pi^{0}$ and $D^{*0} \to D^{0}\gamma$
events in the $D^0\bar{D}^0$ recoil mass spectra without
reconstructing the $\pi^{0}$ or $\gamma$.

To increase the statistics and limit backgrounds, three $D^{0}$ decay
modes with large branching fractions and simple topologies are used,
as shown in Table~\ref{mode}. The corresponding five combinations are
labeled as modes I to V.  Combinations with more than one $\pi^0$ or
more than 6 charged tracks are not used in this analysis.

\begin{table}[!htbp]
\centering
\renewcommand{\arraystretch}{1.3}
\caption{\label{mode}The charmed meson tag modes.}
\begin{tabular}{lll}
\hline
\hline
  Mode &Decay of $D^0$ &Decay of $\bar{D}^0$\\
\hline
I&   $D^{0} \to K^{-}\pi^{+}$                & $\bar{D}^{0} \to K^{+}\pi^{-}$\\
II&  $D^{0} \to K^{-}\pi^{+}$                & $\bar{D}^{0} \to K^{+}\pi^{-}\pi^{0}$\\
III& $D^{0} \to K^{-}\pi^{+}\pi^{0}$         & $\bar{D}^{0} \to K^{+}\pi^{-}$\\
IV&  $D^{0} \to K^{-}\pi^{+}$                & $\bar{D}^{0} \to K^{+}\pi^{-}\pi^{+}\pi^{-}$\\
V&   $D^{0} \to K^{-}\pi^{+}\pi^{+}\pi^{-}$  & $\bar{D}^{0} \to K^{+}\pi^{-}$\\
\hline
\hline
\end{tabular}
\end{table}

To select a good charged track, we require that it must originate
within 10~cm to the interaction point in the beam direction and
1~cm in the plane perpendicular to the beam. In addition, a good
charged track should be within $|\cos\theta|<0.93$, where $\theta$
is its polar angle in the MDC. Information from the TOF and energy
loss ($dE/dx$) measurements in the MDC
are combined to form a probability $P_{\pi}$ ($P_{K}$) with a pion
(kaon) assumption. To identify a pion (kaon), the probability
$P_{\pi}$ ($P_{K}$) is required to be greater than 0.1\%, and
$P_\pi>P_K$ ($P_K>P_\pi$). In modes I-III, one oppositely charged
kaon pair and one oppositely charged pion pair are required in the
final state; while in modes IV and V, one oppositely charged kaon
pair and two oppositely charged pion pairs are required.

Photons, which are reconstructed from isolated showers in the EMC, are
required to be at least 20 degrees away from charged tracks and to
have energy greater than 25~MeV in the barrel EMC or 50~MeV in the
end-cap EMC. To suppress electronic noise and energy deposits
unrelated to the signal event, the EMC time~($t$) of the photon
candidate should be coincident with the collision event time, namely 0
$\leq t \leq$ 700~ns. We require at least two good photons
in modes II and III.

\begin{figure*}[!htbp]
  \centering
  \subfigure[]{
  \label{com.1}
  \includegraphics[width=0.43\textwidth]{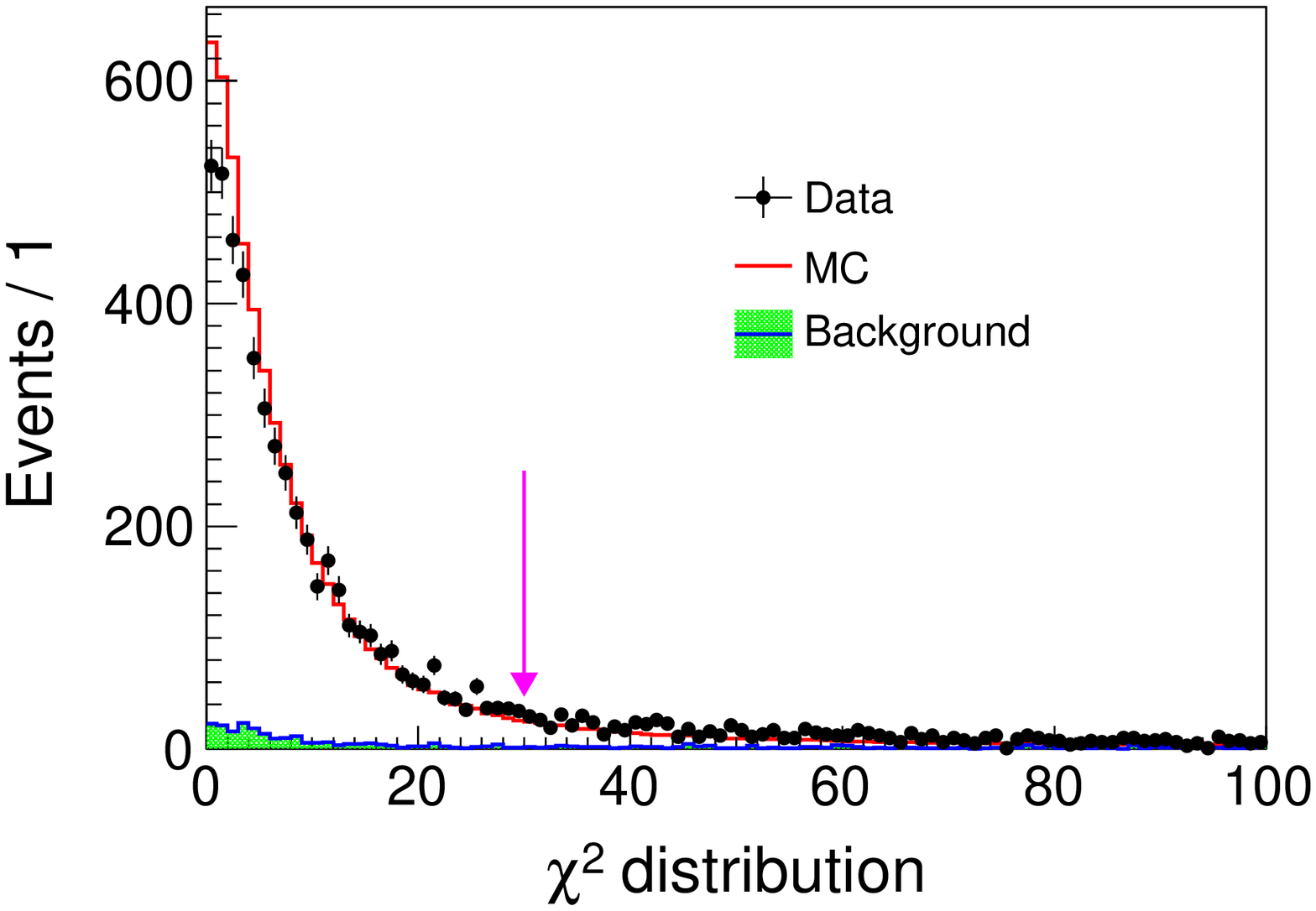}}
  \subfigure[]{
  \label{com.2}
  \includegraphics[width=0.43\textwidth]{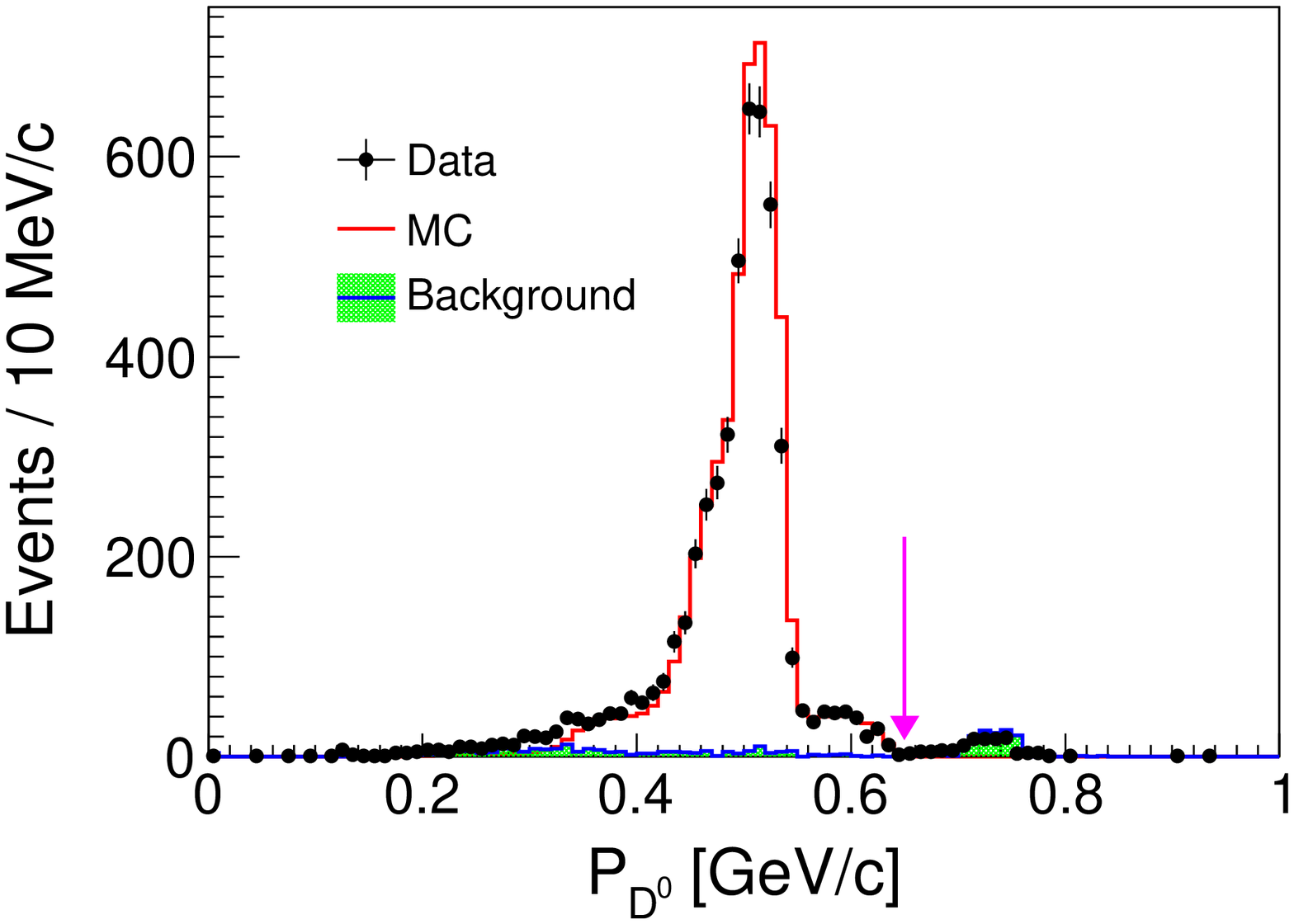}}
  \subfigure[]{
  \label{com.3}
  \includegraphics[width=0.43\textwidth]{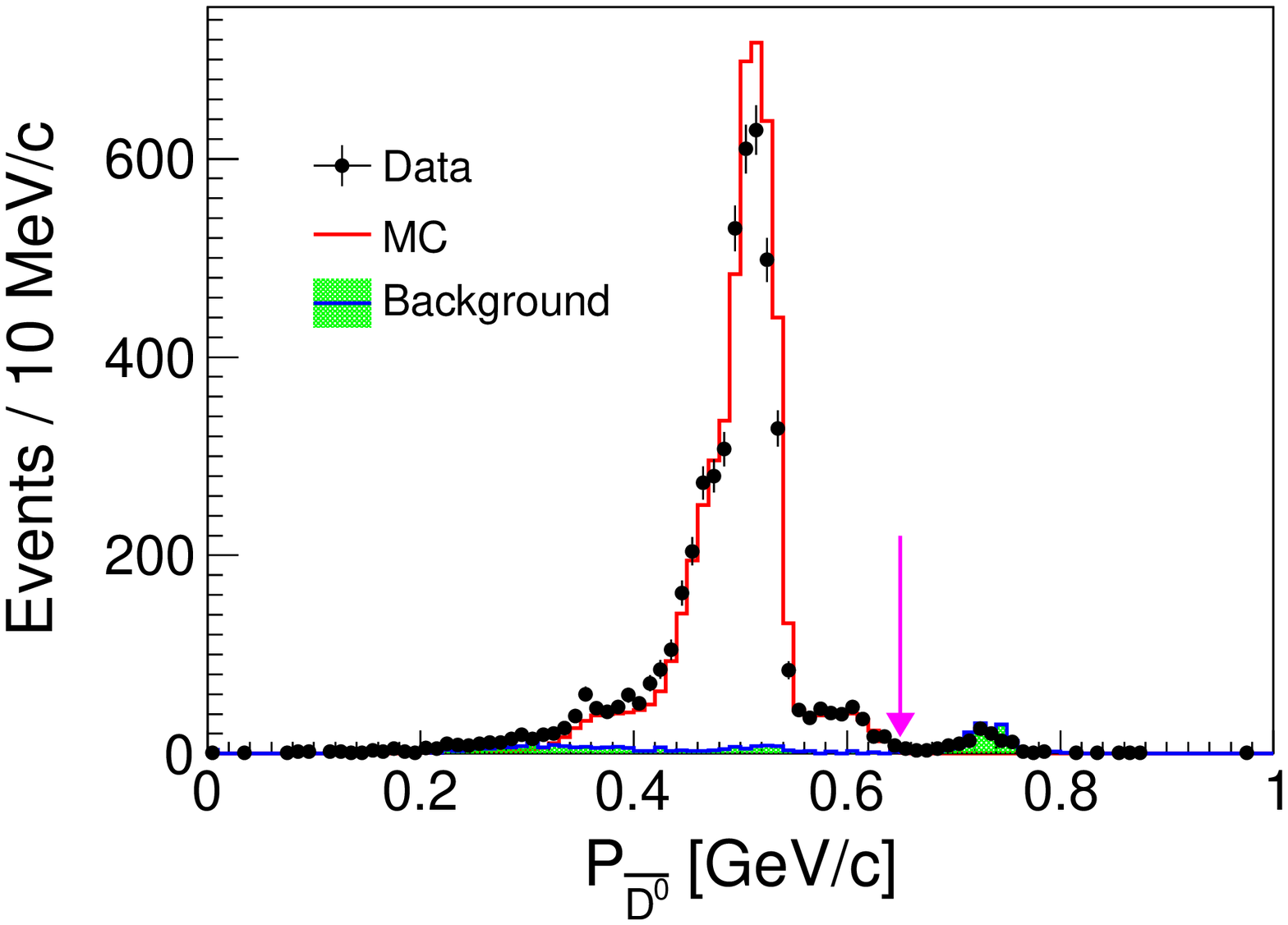}}
  \subfigure[]{
  \label{com.4}
  \includegraphics[width=0.43\textwidth]{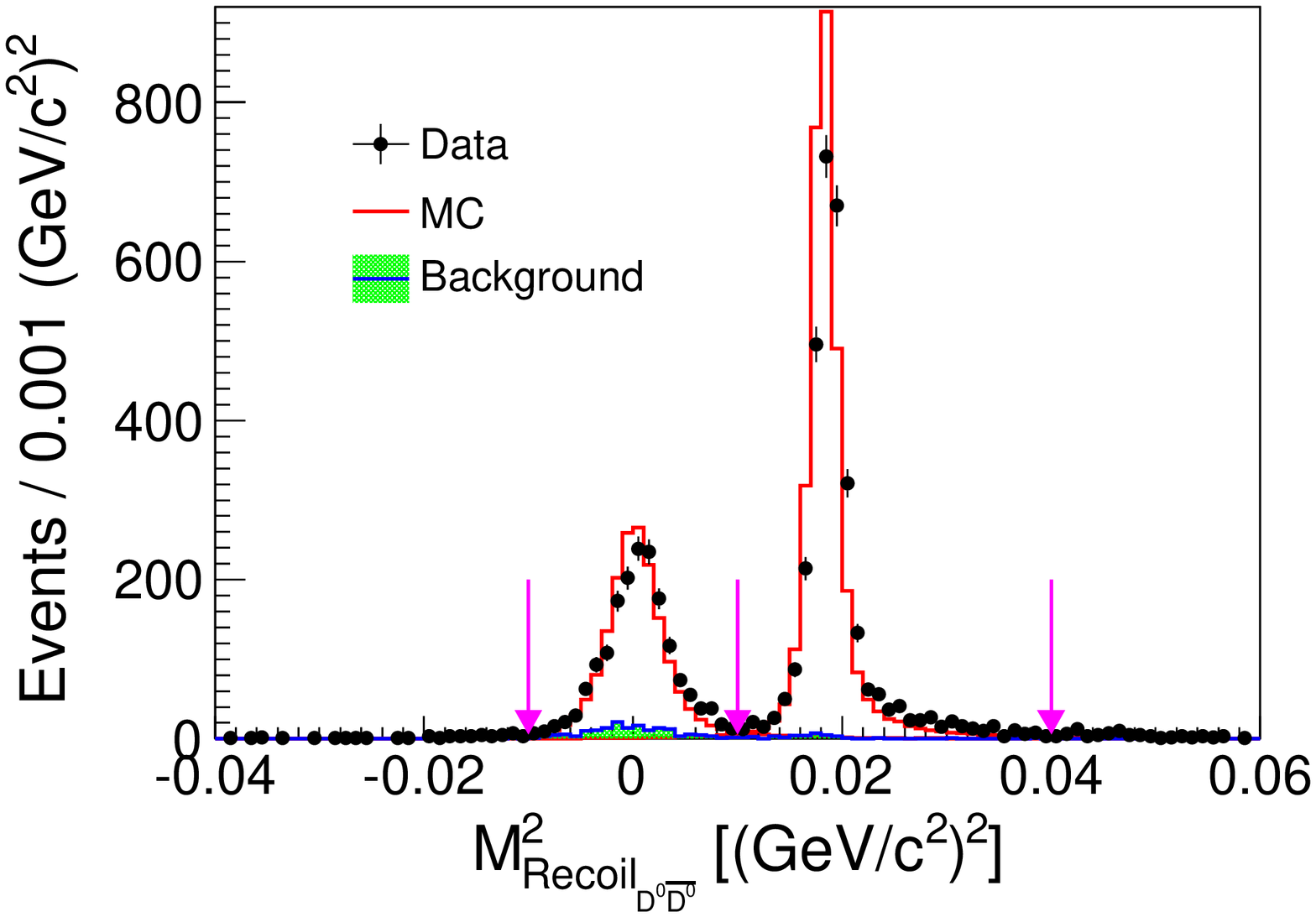}}
\caption{Comparisons between data and MC simulation, summing the
  five modes listed in Table~\ref{mode}: (a) the $\chi^{2}$ distribution, (b) the momentum of
  $D^{0}$, (c) the momentum of $\bar{D}^0$, and (d) the square of the
  $D^{0}\bar{D}^{0}$ recoil mass.  Dots with error bars are data, the
  open red histograms are MC simulations, and the filled green histograms are background
  events from the inclusive MC sample. The signal MCs are normalized
  to data according to the number of events, and background events
  from inclusive MC sample are normalized to data by luminosity.}
  \label{com}
  \end{figure*}

In order to improve the resolution of the $D^{0}\bar{D}^{0}$ recoil
mass, a kinematic fit is performed with the $D^0$ and $\bar{D}^0$ candidates
constrained to the nominal $D^0$ mass~\cite{pdg}. In modes II and III, after
requiring the invariant mass of the two photons be within $\pm
15$~MeV/$c^2$ of the nominal $\pi^0$ mass, a $\pi^{0}$ mass constraint
is also included in the fit. The total $\chi^{2}$ is calculated for
the fit, and when there is more than one $D^{0}\bar{D}^{0}$
combinations satisfying the selection criteria above, the one with the
least total $\chi^{2}$ is selected. Figure~\ref{com} shows comparisons
of some interesting distributions between MC simulation and data after
applying the selection criteria above. Reasonable agreement between
data and MC simulation is observed, and the differences are considered
in the systematic uncertainty estimation. Figure~\ref{com.1} shows the
total $\chi^{2}$ distribution; $\chi^{2}$ less than 30 is required to
increase the purity of the signal. Figures~\ref{com.2} and~\ref{com.3}
show the distributions of $D^{0}$ momentum and $\bar{D^{0}}$ momentum
in the $e^+e^-$ center-of-mass system. The small peaks at
0.75~GeV/$c$ are from direct $e^+e^- \to D^{0}\bar{D}^{0}$
production. To suppress such background events, we require that the
momenta of both $D^{0}$ and $\bar{D}^0$ to be less than
0.65~GeV/$c$. Another source of background events is ISR
production of $\psi(3770)$ with subsequent decay $\psi(3770) \to D^{0}\bar{D}^{0}$,
the number of which is obtained from MC simulation.
As shown in Fig.~\ref{com.4}, the right and left peaks
in the distribution of the square of the $D^{0}\bar{D}^{0}$ recoil
mass correspond to $D^{*0} \to D^{0}\pi^{0}$ and $D^{*0}
\to D^{0}\gamma$ events respectively; the respective signal regions are
defined by $[0.01, 0.04]$ and $[-0.01, 0.01]~({\rm GeV}/c^2)^2$ in the
further analysis.

\section{Branching fractions}

We calculate the branching fraction of $D^{*0}\to D^{0} \pi^0$
using $\BR(D^{*0}\to D^{0} \pi^0) = \frac {N_{\pi^0}^{\rm
prod}}{N_{\gamma}^{\rm prod}+N_{\pi^0}^{\rm prod}}$, where
$N_{\gamma}^{\rm prod}$ and $N_{\pi^0}^{\rm prod}$ are the numbers
of produced $D^{*0}\to D^{0}\gamma$ and $D^{*0}\to D^{0}\pi^{0}$
events, respectively, which are obtained by solving the following
equations
\begin{equation}
 \left( \begin{array}{l}
N_{\pi^0}^{\rm obs}-N_{\pi^0}^{\rm bkg} \\
N_{\gamma}^{\rm obs}-N_{\gamma}^{\rm bkg} \end{array}
 \right) =
 \left( \begin{array}{cc}
\epsilon_{\pi^0\pi^0}  & \epsilon_{\gamma\pi^0}  \\
\epsilon_{\pi^0\gamma} & \epsilon_{\gamma\gamma} \end{array}
 \right)
 \left( \begin{array}{l}
N_{\pi^0}^{\rm prod}  \\
N_{\gamma}^{\rm prod} \end{array}
 \right),
 \end{equation}
where $N_{i}^{\rm obs}$ and $N_{i}^{\rm bkg}$ are the number of
selected events in data and the number of background events
estimated from MC simulation in the $D^{*0}\to D^0+i$ mode,
respectively; $\epsilon_{ij}$ is the efficiency of selecting the
generated $D^{*0}\to D^0+i$ events as $D^{*0} \to  D^0+j$,
determined from MC simulation. Here, $i$ and $j$ denote $\pi^0$ or
$\gamma$. In the simulation, all decay channels of the
$\pi^0$ from $D^{*0}$ decays are taken into account.

\begin{figure}[!htbp]
\centerline{\hbox{ \epsfig{file=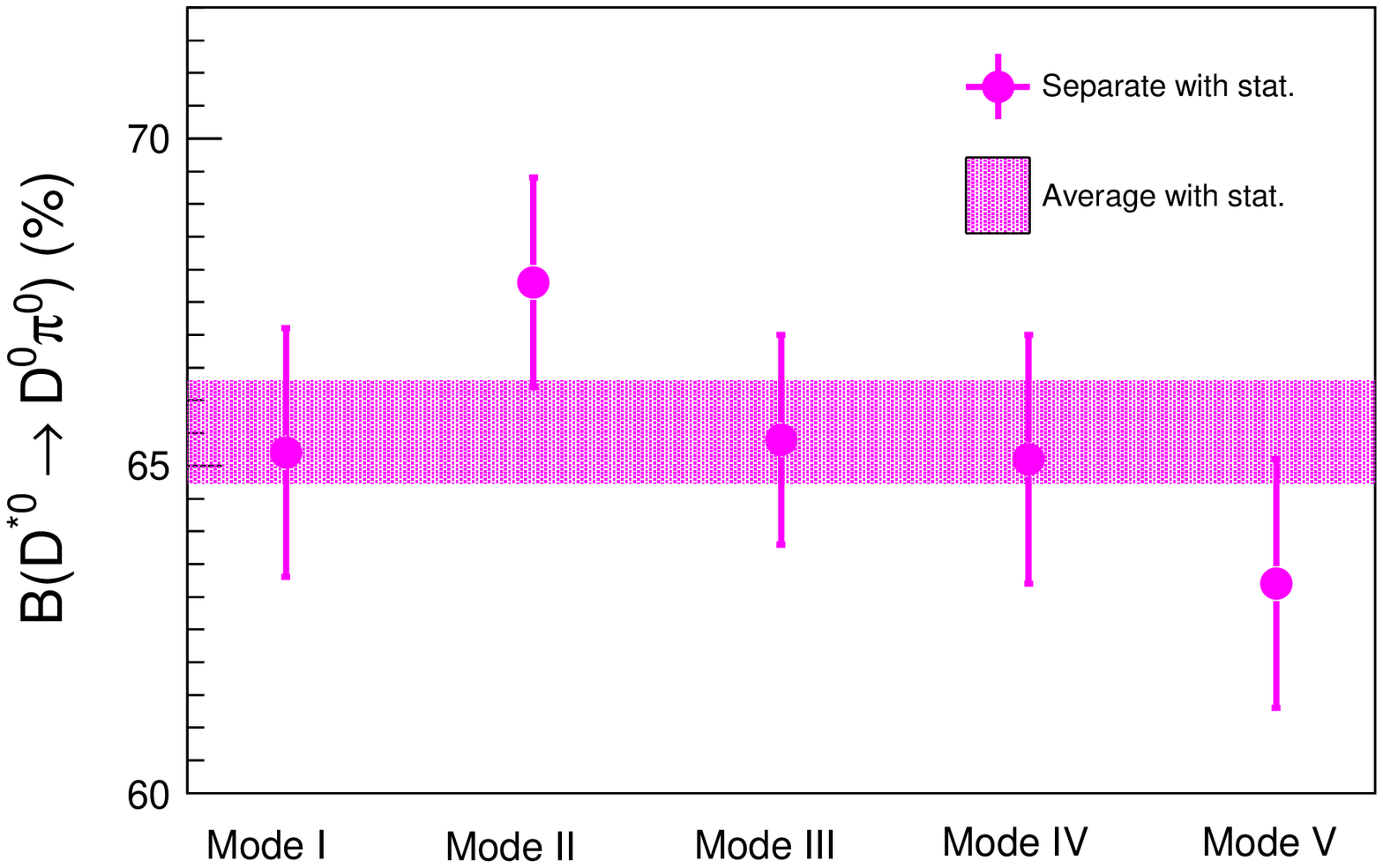,width=8.6cm}}}
\caption{The branching fraction of $D^{*0}\to D^{0}\pi^{0}$. The
  dots with error bars are the results from the five modes; the band
  represents the weighted average. Only statistical uncertainties are
  included.} \label{com_result}
\end{figure}

The numbers used in the calculation and the measured branching
fractions are listed in Table~\ref{pre}. For mode II and III, the final state used to reconstruct the charm meson contains a $\pi^{0}$, so the efficiency for $D^{*0}\to D^{0}\pi^{0}$ will be higher when the $\pi^{0}$ outside the charm meson is misidentified as the $\pi^{0}$ from charm meson decays; for the other three modes, the efficiency difference is caused by the dividing line, this can be illustrated by the fact that $\epsilon_{\pi^0\pi^0}$+$\epsilon_{\pi^0\gamma}$ almost equals to $\epsilon_{\gamma\gamma}$+$\epsilon_{\gamma\pi^0}$. The results from each mode
and their weighted average are shown in Fig.~\ref{com_result}; the
goodness of the fit determined with respect to the weighted average is
$\rm \chi^2/n.d.f.=3.6/4$, which means that the results from these
five modes are consistent with each other.  Here $\rm n.d.f.$ is the number of
degrees of freedom.
The combined result~($\BR(D^{*0}\to D^{0} \pi^0)=65.7\pm0.8\%$), which is calculated by directly summing the
number of events for the five modes together, is consistent with the
weighted average~($\BR(D^{*0}\to D^{0} \pi^0)=65.5\pm0.8\%$). The weighted average is taken as the nominal result.
A cross check is performed by fitting the square of
the $D^{0}\bar{D}^{0}$ recoil mass from data with the MC simulated
signal shapes, and the results agree well with those in
Table~\ref{pre}.

\begin{table*}[!htbp]
\centering
\renewcommand{\arraystretch}{1.3}
\caption{\label{pre} Numbers used for the calculation of the branching
  fractions and the results. $\BR_{\pi^0}$ and $\BR_\gamma$ are the
  the branching fractions of $D^{*0}\to D^{0}\pi^{0}$ and $D^{*0}
  \to D^{0}\gamma$, respectively. ``Combined'' is the result obtained by summing the number of events for the five modes together; ``weighted'' averaged is the result from averaging the results from the five modes by taking the error in each mode as weighted factor. The uncertainties are statistical only.}
\begin{tabular}{lcccccccccc}
 \hline
 \hline
  Mode & $N^{\rm obs}_{\pi^0}$ & $N^{\rm obs}_{\gamma}$ &
  $N^{\rm bkg}_{\pi^0}$ & $N^{\rm bkg}_{\gamma}$ &
  $\epsilon_{\pi^0\pi^0}$ (\%) & $\epsilon_{\gamma\gamma}$ (\%) &
  $\epsilon_{\pi^0\gamma}$ (\%) & $\epsilon_{\gamma\pi^0}$ (\%) &
  $\BR_{\pi^{0}}$ (\%)&$\BR_{\gamma}$ (\%) \\ \hline
 I       &504$\pm$23  &281$\pm$17 &4$\pm$2&24$\pm$5  &36.19&35.22&0.11&0.99&65.2$\pm$1.9&34.8$\pm$1.9\\
 II      &831$\pm$29  &419$\pm$21 &5$\pm$2&36$\pm$6  &15.54&14.46&0.47&0.65&67.8$\pm$1.6&32.2$\pm$1.6\\
 III     &780$\pm$28  &441$\pm$21 &6$\pm$3&38$\pm$6  &15.37&14.60&0.43&0.51&65.4$\pm$1.6&34.6$\pm$1.6\\
 IV      &538$\pm$24  &301$\pm$18 &10$\pm$3&30$\pm$6 &19.04&18.34&0.09&0.51&65.1$\pm$1.9&34.9$\pm$1.9\\
 V       &518$\pm$23  &320$\pm$18 &11$\pm$3&35$\pm$6 &19.05&18.48&0.11&0.53&63.2$\pm$1.9&36.8$\pm$1.9\\
 \hline
 Combined  &&&&&&&&& 65.7$\pm$0.8 & 34.3$\pm$0.8 \\
 Weighted average  &&&&&&&&& 65.5$\pm$0.8 & 34.5$\pm$0.8 \\
 \hline\hline
\end{tabular}
\end{table*}

\section{Systematic uncertainties}

In this analysis, the reconstruction of the photon or the $\pi^{0}$ is
not required. The branching fractions are obtained from the ratio of
the numbers of events in the ranges defined above, so many of the
systematic uncertainties related to the $D^{0}\bar{D}^{0}$
reconstruction, such as the tracking efficiencies, particle
identification efficiencies, etc., cancel.

We use $M^2_{Recoil_{D^0 \bar{D^0}}} = 0.01~({\rm GeV}/c^{2})^{2}$ as the
dividing line between $D^{*0}\to D^{0}\pi^{0}$ and $D^{*0}\to
D^{0}\gamma$, as shown in Fig.~\ref{com.4}.  The systematic
uncertainty due to this selection is estimated by comparing the branching fractions
    via changing this requirement from 0.01 to
0.008 or 0.012~$({\rm GeV}/c^{2})^{2}$.

The $D^{*0}\to D^{0}\pi^{0}$ and $D^{*0}\to D^{0}\gamma$ signal
regions in the $D^{0}\bar{D^{0}}$ recoil mass squared spectrum are in
the combined range of $[-0.01, 0.04]~({\rm GeV}/c^{2})^{2}$; the
associated systematic uncertainty is estimated by removing this
requirement.

The corrected track parameters are used in the nominal MC simulation according to the procedure
described in Ref.~\cite{Ablikim:2012pg}, and the difference in the
branching fractions measured with and without this correction are
taken as the systematic uncertainty caused by the requirement on the
$\chi^{2}$ of the kinematic fit.

The fraction of events with final state radiation~(FSR)
photons from charged pions in data is found to be 20\% higher than
that in MC simulation~\cite{BESIII:2011ac}, and the associated systematic
uncertainty is estimated by enlarging the ratio of FSR events in MC
simulation by a factor of $1.2^{X}$, where $X$ is the number of charged pion in the final state,
and taking the difference in the final result as systematic uncertainty.

The number of background events is calculated from the inclusive MC
sample; the corresponding systematic uncertainty is estimated from the uncertainties of
cross sections used in generating this sample.  The dominant
background events are from open charm processes and ISR production of
$\psi(3770)$ with subsequent $\psi(3770)\to D^{0}\bar{D}^0$. The cross section for
open charm processes is 7.1~nb, with an uncertainty of 0.31~nb or
about 5\%~\cite{CroninHennessy:2008yi}. The cross section for ISR production of
$\psi(3770)$ is 0.114~nb, with an uncertainty of 0.011~nb or about 9\%
which is calculated by varying $\Gamma_{ee}$ and $\Gamma_{\rm total}$
of $\psi(3770)$ by $1\sigma$. The systematic uncertainty related to
the number of background events is conservatively estimated by changing the background level in
Table~\ref{pre} by 10\%~(larger than 5\% and 9\% mentioned above).

The efficiency in Table~\ref{pre} is calculated using 200,000
signal MC events for each mode, but only the ratio
of the efficiencies for $D^{*0}\to D^{0}\pi^{0}$ and
$D^{*0}\to D^{0}\gamma$ is needed in the branching fraction
measurement. The systematic error caused by the
statistical uncertainty of the MC samples is estimated by varying the
efficiency for $D^{*0}\to D^{0}\gamma$ by $1\sigma$ of its
statistical uncertainty, and the difference of the branching
fraction is taken as the systematic uncertainty.

Other possible systematic uncertainty sources, such as from the
simulation of ISR, the requirement on the charmed meson momentum, and the
tracking efficiency difference caused by the tiny phase space
difference between the two decay modes of $D^{*0}$, are investigated
and are negligible.

The summary of the systematic uncertainties considered is shown in
Table~\ref{sys}. Assuming the systematic uncertainties from the
different sources are independent, the total systematic uncertainty is
found to be 0.5\% by adding all the sources in quadrature.

\begin{table}[!htbp]
\centering
\renewcommand{\arraystretch}{1.3}
\caption{\label{sys}The summary of the absolute systematic
uncertainties in $\BR(D^{*0}\to D^0\pi^0)$ and $\BR(D^{*0}\to
D^0\gamma)$.}
\begin{tabular}{lllllllllll}
\hline\hline
 Source     &(\%)\\
\hline
 Dividing line between $D^{*0} \to D^0 \pi^0$ and  $D^{*0} \to D^0 \gamma$                                           &0.2\\
 Choice of signal regions                              &0.2\\
 Kinematic fit                                         &0.2\\
 FSR simulation                                        &0.1\\
 Background                                            &0.2\\
 Statistics of MC samples                              &0.2\\
\hline
 Sum                                                   &0.5\\
\hline\hline
\end{tabular}
\end{table}

\section{Summary}

By assuming that there are only two modes of $D^{*0}$, we measure the branching fractions of $D^{*0}$ to be $\BR(D^{*0}\to
D^0\pi^0)=(65.5\pm 0.8\pm 0.5)\%$ and $\BR(D^{*0}\to
D^0\gamma)=(34.5\pm 0.8\pm 0.5)\%$, where the first uncertainties
are statistical and the second ones are systematic. It should be
noted that both the statistical and the systematic uncertainties
of these two branching fractions are fully anti-correlated.
Taking the correlations into account, the branching ratio
$\BR(D^{*0}\to D^0\pi^0)/\BR(D^{*0} \to D^0\gamma) = 1.90\pm
0.07\pm 0.05$ is obtained. This ratio does not depend on any
assumptions in the $D^{*0}$ decays, so it can be used in
calculating the $D^{*0}$ decay branching fractions if more decay
modes are discovered.

\begin{figure}[!htbp]
\centerline{\hbox{ \epsfig{file=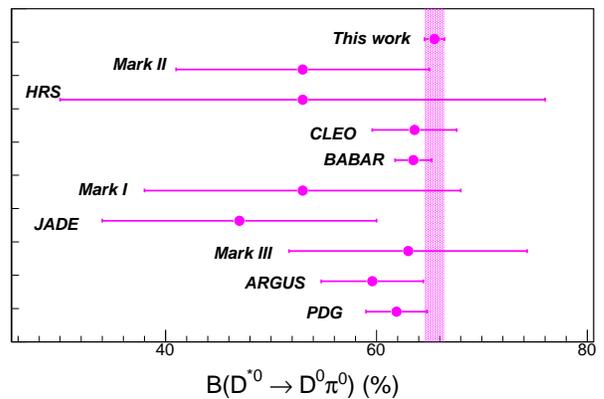,width=8.6cm}}}
\caption{Comparison of the branching fraction of $D^{*0}\to
D^{0}\pi^{0}$ from this work and from previous experiments. Dots
with error bars are results from different experiments, and the
band is the result from this work with both statistical and
systematic uncertainties.} \label{com_all}
\end{figure}

Figure~\ref{com_all} shows a comparison of the measured branching
fraction of $D^{*0} \to D^{0}\pi^{0}$ with other experiments and the
world average value~\cite{pdg}. Our measurement is consistent with the
previous ones within about 1$\sigma$ but with much better
precision. These much improved results can be used to update the
parameters in the effective models mentioned above, such as the mass of the charm
quark~\cite{Eichten:1979ms, Aliev:1994nq}, the effective coupling
constant~\cite{Cheng:1993gc}, and the magnetic moment of the charm
quark~\cite{Miller:1988tz}. With these new results as input, the
uncertainty in the semileptonic decay branching fraction of
$B^{\pm}$~\cite{Bozek:2010xy} can be reduced, thus leading to a
tighter constraint on the standard model~(SM) and its extensions.

\begin{acknowledgements}
  The BESIII collaboration thanks the staff of BEPCII and the IHEP
  computing center for their strong support. This work is supported in
  part by National Key Basic Research Program of China under Contract
  No.~2015CB856700; Joint Funds of the National Natural Science
  Foundation of China under Contracts Nos.~11079008, 11179007,
  U1232201, U1332201; National Natural Science Foundation of China
  (NSFC) under Contracts Nos.~10935007, 11121092, 11125525, 11235011,
  11322544, 11335008; the Chinese Academy of Sciences (CAS) Large-Scale
  Scientific Facility Program; CAS under Contracts Nos.~KJCX2-YW-N29,
  KJCX2-YW-N45; 100 Talents Program of CAS; INPAC and Shanghai Key
  Laboratory for Particle Physics and Cosmology; German Research
  Foundation DFG under Contract No.~Collaborative Research Center
  CRC-1044; Istituto Nazionale di Fisica Nucleare, Italy; Ministry of
  Development of Turkey under Contract No.~DPT2006K-120470; Russian
  Foundation for Basic Research under Contract No.~14-07-91152;
  U.S. Department of Energy under Contracts Nos.~DE-FG02-04ER41291,
  DE-FG02-05ER41374, DE-FG02-94ER40823, DESC0010118; U.S. National
  Science Foundation; University of Groningen (RuG) and the
  Helmholtzzentrum fuer Schwerionenforschung GmbH (GSI), Darmstadt; WCU
  Program of National Research Foundation of Korea under Contract
  No.~R32-2008-000-10155-0.

\end{acknowledgements}

\end{document}

%% file: authors_dec2014.tex
\author{
  \begin{small}
    \begin{center}
      M.~Ablikim$^{1}$, M.~N.~Achasov$^{8,a}$, X.~C.~Ai$^{1}$,
      O.~Albayrak$^{4}$, M.~Albrecht$^{3}$, D.~J.~Ambrose$^{43}$,
      A.~Amoroso$^{47A,47C}$, F.~F.~An$^{1}$, Q.~An$^{44}$,
      J.~Z.~Bai$^{1}$
% ...
      \\
      \vspace{0.2cm}
      (BESIII Collaboration)\\
      \vspace{0.2cm} {\it
        $^{1}$ Institute of High Energy Physics, Beijing 100049, People's Republic of China\\
% ...
$^{3}$ Bochum Ruhr-University, D-44780 Bochum, Germany\\
        $^{4}$ Carnegie Mellon University, Pittsburgh, Pennsylvania 15213, USA\\
% ...
$^{8}$ G.I. Budker Institute of Nuclear Physics SB RAS (BINP), Novosibirsk 630090, Russia\\
% ...
$^{43}$ University of Rochester, Rochester, New York 14627, USA\\
        $^{44}$ University of Science and Technology of China, Hefei 230026, People's Republic of China\\
% ...
$^{47}$ (A)University of Turin, I-10125, Turin, Italy; (B)University of Eastern Piedmont, I-15121, Alessandria, Italy; (C)INFN, I-10125, Turin, Italy\\
% ...
$^{a}$ Also at the Novosibirsk State University, Novosibirsk, 630090, Russia\\
% ...
      }\end{center}
    \vspace{0.4cm}
  \end{small}
}

\affiliation{}